\documentclass[aps,prd,twocolumn,groupedaddress,showpacs]{revtex4-1}
\usepackage{epsf,epsfig,graphicx}
\usepackage{amsmath}
\usepackage{amssymb}
\usepackage{tikz}

\def\edth{\;\raise1.0pt\hbox{$'$}\hskip-6pt\partial\;}
\def\baredth{\;\overline{\raise1.0pt\hbox{$'$}\hskip-6pt
\partial}\;}
\def\gsim{~\rlap{$>$}{\lower 1.0ex\hbox{$\sim$}}}

\newcommand{\be}{\begin{equation}}
\newcommand{\ee}{\end{equation}}
\newcommand{\bw}{\begin{widetext}}
\newcommand{\ew}{\end{widetext}}

\newcommand{\intinf}{\int_{-\infty}^{\infty}}
\newcommand{\suml}{\sum_{l=0}^{\infty}}
\newcommand{\summ}{\sum_{m=-\ell}^{\ell}}

\usepackage{xcolor}

\AtBeginDocument{\renewcommand{\d}{\mathbf{d}}}

\begin{document}

\title{Observation of polarized stochastic gravitational-wave background in pulsar-timing-array experiments}

\author{Yu-Kuang Chu$^{1}$, Guo-Chin Liu$^{2}$, and Kin-Wang Ng$^{1,3}$}

\affiliation{
$^1$Institute of Physics, Academia Sinica, Taipei 11529, Taiwan\\
$^2$Department of Physics, Tamkang University, Tamsui, New Taipei City 25137, Taiwan\\
$^3$Institute of Astronomy and Astrophysics, Academia Sinica, Taipei 11529, Taiwan
}

\vspace*{0.6 cm}
\date{\today}
\vspace*{1.2 cm}

\begin{abstract}
We study the observation of polarized stochastic gravitational-wave background (SGWB) in pulsar-timing-array experiments. The time residual for an observed pulsar is formulated as a line-of-sight integral that incorporates the effects of the pulsar term, from which we construct the correlation function of the time residual between a pair of pulsars in terms of the overlap reduction functions (ORFs) for the SGWB intensity and polarization anisotropies. Our formulation provides a numerical scheme for computing the ORFs for high multipole moments and the lowest-moment ORFs for the SGWB linear polarization are worked out for the first time.
\end{abstract}

\maketitle

\section{Introduction}

The existence of gravitational waves (GWs), considered as ripples in space-time fabric, 
was predicted in Einstein's theory of general relativity in 1916. 
It was only until recently that GWs emitted by a binary black hole merger were detected by the LIGO experiment~\cite{ligo}.
Since the first detection, many GW events from compact binary coalescences have been observed in Advanced LIGO and Advanced Virgo O2 and O3 observing runs~\cite{ligo2019}. This opens up a new era of GW astronomy and cosmology. Upcoming and future GW experiments such as KAGRA, GEO600, Einstein Telescope, Cosmic Explorer, LISA, DECIGO, Taiji, and TianQin, will bring us a precision science in GW observation~\cite{ligo2050}.

Stochastic gravitational wave background (SGWB) is one of the main goals in GW experiments. There have been many studies on potential astrophysical and cosmological sources for SGWB such as distant compact binary coalescences, early-time phase transitions, cosmic string or defect networks, second-order primordial scalar perturbations, and inflationary GWs~\cite{romano}. GWs are very weakly interacting, so they decouple from matter at the time of production and then travel to us almost without being scattered. At the present, they remain as a SGWB that carries initial conditions of the production processes in the early Universe. 

In general, the SGWB can be anisotropic and polarized. For examples, helical GWs can be produced in axion inflation models, leading to a net circular polarization~\cite{alexander,satoh,sorbo,crowder2013}. Linear polarization can be generated through diffusion by compact astrophysical objects, with an amount suppressed by a factor of at least $10^{-4}$ with respect to the intensity anisotropies; however, it can be enhanced if dark matter is dominated with sub-solar-mass primordial black holes~\cite{cusin}. Furthermore, the directionality dependence of the SGWB have been recently explored~\cite{bartolo,pitrou}.

The method adopted in current GW interferometry such as the LIGO-Virgo experiment for detecting the intensity and polarization anisotropies or the Stokes parameters of the SGWB is to correlate the responses of a pair of detectors to the GW strain amplitude. The correlation allows us to filter out detector noises and increase the signal-to-noise ratio~\cite{romano}. The on-going and future GW interferometry experiments have sensitivities for detecting GWs at frequencies ranging from kilohertz to millihertz.

Another method to search for SGWB anisotropies is through their gravitational effects on the arrival times of radio pulses from millisecond pulsars~\cite{romano}. In the pulsar-timing observation, radio pulses from an array of roughly 100 Galactic millisecond pulsars are being monitored with ground-based radio telescopes. The fluctuations in the redshifts of the radio pulses due to the presence of GWs would allow us to detect longer-wavelength GWs at nanohertz frequencies. Current pulsar timing array (PTA) experiments include EPTA~\cite{EPTA},  NANOGrav~\cite{NANO}, and PPTA~\cite{PPTA}. The future SKA radio telescope will have projected sensitivity three to four orders of magnitude better than the current PTAs~\cite{SKA}.
Recently, the NANOGrav Collaboration~\cite{nanograv} has found strong evidence of a stochastic common-spectrum process across 45 millisecond pulsars, alluding to a SGWB with a characteristic strain of $h_c=1.92\times10^{-15}$ 
at a reference frequency of $f_{\rm yr}=1\,{\rm yr^{-1}}\simeq 31.8\,{\rm nHz}$. 
However, they have not found statistically significant evidence that this process has quadrupolar spatial correlations, as expected for the properties in the presence of a SGWB.

The pulsar-timing observation of the SGWB has been studied extensively~\cite{romano}. The fluctuation in the redshift of the radio pulse from a pulsar in the presence of GWs is due to the Shapiro time delay given by the time-varying gravitational potential difference from the pulsar to the Earth. By ignoring the contribution at the pulsar site (the so-called pulsar term), the overlap reduction functions (ORFs) for the SGWB intensity anisotropy have been derived using the two-point correlation method~\cite{anholm,mingar13}. Later, it has been extended to including the SGWB circular polarization anisotropy~\cite{kato16}. Furthermore, the corrections of the ORFs for the SGWB intensity anisotropy by the pulsar term have been discussed~\cite{mingar14}. In this paper, we develop a numerical scheme to compute the ORFs for the SGWB intensity and polarization anisotropies, following the spherical harmonic expansion method adopted in Ref.~\cite{chu21} though in the context of GW interferometry.  Instead of using the Shapiro time delay, we follow Ref.~\cite{ng21} to begin with the Sachs-Wolfe line-of-sight integral and then calculate the two-point correlation function. This scheme allows us to compute the ORFs for high multipole moments order by order and simultaneously incorporates the effects of the pulsar term. 

\section{Polarized SGWB}

In the Minkowskian vacuum, the metric perturbation $h_{ij}$ in the transverse traceless gauge depicts GWs propagating at the speed of light $c=\omega/k$. At a given spacetime point $(t,\vec{x})$, it can be expanded in terms of its Fourier modes:
\be
\label{eq:planwave}
h_{ij}(t,\vec{x}) = \sum_{A}\intinf \d f \int_{S^2} \d\hat{k} \;
    h_A(f,\hat{k}) \mathbf{e}^A_{ij}(\hat{k})
    e^{-2 \pi i f (t - \hat{k}\cdot \vec{x}/c)}\,,
\ee
where $A$ stands for the polarization or the helicity of GWs described by the corresponding basis tensors $\mathbf{e}^A_{ij}(\hat{k})$, which are transverse to the direction of the wave propagation denoted by $\hat{k}$. Since $h_{ij}$ is real, its Fourier components are not fully independent with each other. For our application, we require those Fourier components with negative frequencies to be $h_A(-f,\hat{k})=h_A^*(f,\hat{k})$ for all $f\ge 0$. The GWs are considered as stochastic as long as $h_{ij}$ are random fields thus characterized by their ensemble averages. Besides, assuming the probability distribution of the random amplitude $h_{ij}$ be Gaussian, then only the two-point correlation function $\langle h_{ij}(t,\vec{x}_1) h_{ij}(t,\vec{x}_2) \rangle$ is needed to describe its statistical behavior. Furthermore, if the waves are homogeneous, i.e., having translational symmetry, the ensemble average can be evaluated by doing spatial averages. As a result, the two-point correlation function of the Fourier modes should have the following form  
\be
\label{eq:paa}
\langle h_{A}(f,\hat{k}) h^*_{A'}(f',\hat{k}') \rangle
= \delta(f-f') \delta(\hat{k}-\hat{k}')P_{AA'}(f,\hat{k}) \,,
\ee
where the $\delta(f-f')$ arises from the delta function of the magnitude of their 3-momenta $\delta(\vec{k}-\vec{k}')$ and the assumption made in Eq.~(\ref{eq:planwave}) that these waves satisfy the equation of motion in vacuum. Also, the presence of $\delta(f-f')$ implies that the signal is stationary. This is a fairly good approximation during an observing period for a typical experiment.

For GWs coming from the sky direction $-\hat{k}$ with wave vector $\vec{k}$, it is customary to write the polarization basis tensors in terms of the basis vectors in the spherical coordinates:
\begin{align}
    \mathbf{e}^{+}(\hat{k}) &= \hat{\mathbf{e}}_\theta \otimes \hat{\mathbf{e}}_\theta  
                               - \hat{\mathbf{e}}_\phi \otimes \hat{\mathbf{e}}_\phi \,, \nonumber \\
    \mathbf{e}^{\times}(\hat{k}) &= \hat{\mathbf{e}}_\theta \otimes \hat{\mathbf{e}}_\phi  
                                    + \hat{\mathbf{e}}_\phi \otimes \hat{\mathbf{e}}_\theta  \,,
\end{align}
in which $\hat{\mathbf{e}}_\theta$, $\hat{\mathbf{e}}_\phi$, and $\hat{k}$ form a right-handed orthonormal basis.
Also, we can define the complex circular polarization basis tensors as
\begin{align}
    \mathbf{e}_{R} &= \frac{(\mathbf{e}_{+} + i \mathbf{e}_{\times})}{\sqrt{2}} \,,
   &\mathbf{e}_{L} &= \frac{(\mathbf{e}_{+} - i \mathbf{e}_{\times})}{\sqrt{2}} \,,
\end{align}
where $\mathbf{e}_{R}$ stands for the right-handed GW with a positive helicity while $\mathbf{e}_{L}$ stands for the left-handed GW with a negative helicity. The corresponding amplitudes in Eq.~(\ref{eq:planwave}) in the two different bases are related to each other via
\begin{align}
    h_{R} &= \frac{(h_{+} - i h_{\times})}{\sqrt{2}} \,,
   &h_{L} &= \frac{(h_{+} + i h_{\times})}{\sqrt{2}} \,.
\end{align}

Analogous to the case in electromagnetic waves~\cite{book:BornAndWolf}, the coherency matrix $P_{AA'}$ in Eq.~(\ref{eq:paa}) is related to the Stokes parameters, $I$, $Q$, $U$, and $V$ as 
\begin{align}
    I &= \left[ \langle h_R h_R^* \rangle + \langle h_L h_L^* \rangle \right] / 2 \,,\nonumber\\
    Q + iU &=  \langle h_L h_R^*  \rangle \,,\nonumber\\
    Q - iU &=  \langle h_R h_L^*  \rangle \,,\nonumber\\
    V &= \left[ \langle h_R h_R^* \rangle - \langle h_L h_L^* \rangle \right] / 2 \,\label{eq:spIQUV}.
\end{align}
They are functions of the frequency $f$ and the propagation direction $\hat{k}$.
To get some flavor of the meaning of these Stokes parameters, we may take a look at an example for an unpolarized quasimonochromatic GW signal with a constant intensity. It should have a constant $I$ which represents the total intensity regardless of its polarization. We have $V=0$ since the power in the right-handed and the left-handed modes should be identical. Also, because the relative phase between the left-handed and the right-handed modes ($\arg(h_L)-\arg(h_R)$) is random for an unpolarized source, the ensemble average $\langle h_L h_R^* \rangle \sim \langle e^{i(\arg(h_L)-\arg(h_R))} \rangle$ becomes zero, thereby making $Q=U=0$. 

\section{Pulsar timing}
\label{ptiming}

In the pulsar-timing observation, radio pulses from an array of roughly 100 Galactic millisecond pulsars are being monitored with ground-based radio telescopes. The redshift fluctuation of a pulsar in the pointing direction 
$\hat{e}$ on the sky is given by the Sachs-Wolfe effect~\cite{sachs},
\begin{equation}
z(\hat{e})= - {1\over 2}\int_{\eta_e}^{\eta_r}\d\eta\, \hat{e}^i \hat{e}^j \frac{\partial}{\partial\eta}h_{ij}(\eta, \vec x)\,,
\label{rsfze}
\end{equation}
where the lower (upper) limit of integration in the line-of-sight integral represents the point of emission (reception) 
of the radio pulse. The physical distance of the pulsar from the Earth is 
\be
D=c(\eta_r-\eta_e)\,, 
\ee
which is of order $1\,{\rm kpc}$.

The quantity that is actually observed in the pulsar-timing observation is the time residual counted as
\begin{equation}
r(t)=\int_0^t dt'z(t')\,,
\end{equation}
where $t'$ denotes the laboratory time and $t$ is the duration of the observation. 
Using the laboratory time $t'$, we rewrite Eq.~(\ref{rsfze}) as
\begin{equation}
z(t',\hat{e})= - {1\over 2}\int_{t'+\eta_e}^{t'+\eta_r}\d\eta\, d^{ij}\frac{\partial}{\partial\eta}h_{ij}(\eta, \vec x)\,,
\end{equation}
where the detector tensor is
\be
d^{ij}=\hat{e}^i \hat{e}^j\,.
\ee
Then, replacing $\vec{x}$ by $c(\eta_r-\eta)\hat{e}$, we have
\bw
\begin{align}
r(t)&={1\over2}\sum_{A}\intinf \d f \int_{S^2} \d\hat{k} \int_{\eta_e}^{\eta_r}\d\eta\,(1-e^{-2\pi i f t}) \,
    h_A(f,\hat{k}) \,d^{ij} \mathbf{e}^A_{ij}(\hat{k})\,
    e^{-2 \pi i f \eta} e^{2\pi i f (\eta_r-\eta) \hat{k} \cdot \hat{e}} \nonumber \\
    &=\frac{1}{4\pi}\sum_{A}\intinf \frac{\d f}{f} \int_{S^2} \d\hat{k}\, (1-e^{-2\pi i f t})\, e^{2\pi i f\eta_r}
    h_A(f,\hat{k})\, d^{ij} \mathbf{e}^A_{ij}(\hat{k}) \left[1-e^{2\pi i f D (1+\hat{k} \cdot \hat{e})/c}\right]\frac{i}{1+\hat{k} \cdot \hat{e}}\,.
\end{align}
\ew
This is the result for the time residual due to the Shapiro time delay, where the first term in the square brackets is called the Earth term and the second is the pulsar term. The last term has a pole at $\hat{k} \cdot \hat{e}=-1$. For Galactic pulsars in a pulsar-timing experiment with a typical duration of observation, we have $fD\gg 1$, so the pulsar term is a highly oscillatory function in the $\hat{k}$-integral. If the separation angle between any pair of pulsars is large enough, we can safely omit the pulsar term to simplify the $\hat{k}$-integral for obtaining the two-point correlation function~\cite{anholm,mingar13,kato16}. 

In the present work, we retain the Sachs-Wolfe line-of-sight integral. This allows us to avoid the pole term and do the spherical harmonic expansion. In addition, the effect of the pulsar term is now put back in the time integral. Then, using the spherical wave expansion~(\ref{eq:swexpansion}) for the phane wave~(\ref{eq:planwave}), we obtain
\bw
\be
r(t)=2\pi\sum_{A}\intinf \d f \int_{S^2} \d\hat{k} \int_{\eta_e}^{\eta_r}\d\eta\,(1-e^{-2\pi i f t}) 
    h_A(f,\hat{k}) d^{ij} \mathbf{e}^A_{ij}(\hat{k})
    e^{-2 \pi i f \eta} \sum_{LM} i^L j_L[2\pi f (\eta_r-\eta)] Y_{LM}^*(\hat{k}) Y_{LM}(\hat{e})\,.
\ee
\ew

\section{Residual Correlation}

The time-residual correlation between a pair of pulsars $a$ and $b$ is constructed as
\be
\langle r(t_a)r(t_b) \rangle = \int_0^{t_a}dt' \int_0^{t_b}dt{''} \langle z(t',\hat{e}_a) z(t{''},\hat{e}_b)\rangle.
\ee
Using Eq.~(\ref{eq:paa}) and defining $x=2\pi f (\eta_r-\eta)$, this becomes
\bw
\begin{align}
\label{eq:xi2gamma}
\langle r(t_a)r(t_b) \rangle=&
  \intinf \frac{\d f}{(2\pi f)^2}\,(1-e^{-2\pi i f t_a}) (1-e^{2\pi i f t_b})   \sum_{L_1 M_1 L_2 M_2} 
       J_{L_1}(x_a) J^*_{L_2}(x_b)\, \times \nonumber \\
       &
       Y_{L_1 M_1}(\hat{e}_a) Y^*_{L_2 M_2}(\hat{e}_b)\,
       \int_{S^2} \d \hat{k} \; R(f,\hat{k},\hat{e}_a,\hat{e}_b)\, Y_{L_1 M_1}^*(\hat{k}) Y_{L_2 M_2}(\hat{k})\,,
\end{align}
\ew
where $x_a=2\pi f D_a/c$, $x_b=2\pi f D_b/c$, and 
\begin{align}
& J_{L_1}(x_a)=  i^{L_1} 2\pi
   \int_0^{x_a}\d x\,e^{ix} j_{L_1}(x),  \\
& J_{L_2}(x_b)=  i^{L_2} 2\pi
   \int_0^{x_b}\d x\,e^{ix} j_{L_2}(x),  \\
&R(f,\hat{k},\hat{e}_a,\hat{e}_b)=
    \sum_{AA'}
    P_{AA'}(f,\hat{k}) \,d_a^{ij} d_b^{kl}\,
    \mathbf{e}^{A}_{ij}(\hat{k}) \mathbf{e}^{*A'}_{kl}(\hat{k})\,.
\end{align} 

In terms of the Stokes parameters in Eq.~(\ref{eq:spIQUV}) and the definitions,
\begin{align}   
    \mathbb{E}^{I}_{ijkl} (\hat{k}) &= 
    \mathbf{e}^{R}_{ij}(\hat{k}) \mathbf{e}^{*R}_{kl}(\hat{k}) + 
    \mathbf{e}^{L}_{ij}(\hat{k}) \mathbf{e}^{*L}_{kl}(\hat{k}) \,,
    \nonumber \\ 
    \mathbb{E}^{V}_{ijkl} (\hat{k}) &= 
    \mathbf{e}^{R}_{ij}(\hat{k}) \mathbf{e}^{*R}_{kl}(\hat{k}) - 
    \mathbf{e}^{L}_{ij}(\hat{k}) \mathbf{e}^{*L}_{kl}(\hat{k}) \,,
    \nonumber \\ 
    \mathbb{E}^{Q+iU}_{ijkl} (\hat{k}) &= 
    \mathbf{e}^{L}_{ij}(\hat{k}) \mathbf{e}^{*R}_{kl}(\hat{k}) \,,
    \nonumber \\ 
    \mathbb{E}^{Q-iU}_{ijkl} (\hat{k}) &= 
    \mathbf{e}^{R}_{ij}(\hat{k}) \mathbf{e}^{*L}_{kl}(\hat{k}) \,,
\end{align}

denoting $\mathbb{D}$ as the direct product of two detector tensors
\be
{}^{ijkl}\mathbb{D}_{ab}=d_a^{ij} d_b^{kl} \,,
\ee
the antenna pattern for signal reception can be rewritten as
\be
 R(f,\hat{k},\hat{e}_a,\hat{e}_b)=
\sum_{X=\{I,V,Q\pm iU\}}
 X(f,\hat{k}) \;
{}^{ijkl}\mathbb{D}_{ab} \;
 \mathbb{E}^{I}_{ijkl} (\hat{k}) \,.
\ee
We further expand the Stokes parameters and the polarization basis tensors in terms of ordinary and spin-weighted spherical harmonics as
\begin{align}
    I(f,\hat{k}) &= \sum_{\ell m}I_{\ell m}(f) \; Y_{\ell m}(\hat{k}) \,,\nonumber \\
    V(f,\hat{k}) &= \sum_{\ell m}V_{\ell m}(f) \; Y_{\ell m}(\hat{k}) \,,\nonumber \\
     (Q+iU)(f,\hat{k}) &= \sum_{\ell m}(Q+iU)_{\ell m}(f) \; _{+4}Y_{\ell m}(\hat{k}) \,,\nonumber \\
    (Q-iU)(f,\hat{k}) &= \sum_{\ell m}(Q-iU)_{\ell m}(f) \; _{-4}Y_{\ell m}(\hat{k}) \,,
\end{align}
and
\begin{align}
    \mathbb{E}^{I}_{ijkl} (\hat{k}) 
    &=\sum_{\ell_e m_e} {}_{ijkl}\mathbb{E}^{I}_{\ell_e m_e} Y_{\ell_e m_e}(\hat{k}) \,, \nonumber
   \\
    \mathbb{E}^{V}_{ijkl} (\hat{k}) 
    &=\sum_{\ell_e m_e} {}_{ijkl}\mathbb{E}^{V}_{\ell_e m_e} Y_{\ell_e m_e}(\hat{k}) \,, \nonumber
   \\
    \mathbb{E}^{Q+iU}_{ijkl} (\hat{k}) 
    &=\sum_{\ell_e m_e} {}_{ijkl}\mathbb{E}^{Q+iU}_{\ell_e m_e} {}_{-4}Y_{\ell_e m_e}(\hat{k}) \,, \nonumber
   \\
    \mathbb{E}^{Q-iU}_{ijkl} (\hat{k}) 
    &=\sum_{\ell_e m_e} {}_{ijkl}\mathbb{E}^{Q-iU}_{\ell_e m_e} {}_{+4}Y_{\ell_e m_e}(\hat{k}) \,,
\end{align}
where the specific combinations, $Q\pm iU$, make them become spin $\pm4$ objects so that we can expand them nicely by the corresponding spin-weighted spherical harmonics. A brief introduction to the spin-weighted spherical harmonics is found in Appendix~\ref{sec:spinweight}. In Appendix~\ref{sec:polartensor}, we have given the multipole moments of the polarization tensors $\mathbb{E}_{ijkl}$ for each Stokes parameter.

Hence we can express the time-residual correlation in the following form,
\bw
\be
\langle r(t_a)r(t_b) \rangle=
  \intinf \frac{\d f}{(2\pi f)^2}\,(1-e^{-2\pi i f t_a}) (1-e^{2\pi i f t_b}) 
   \sum_{X=\{I,V,Q\pm iU\}}
   \sum_{\ell m}  X_{\ell m}(f) \gamma_{\ell m}^{I}(x_a,x_b,\hat{e}_a,\hat{e}_b)\,,
\ee
where the overlap reduction functions (ORFs) are given by
\begin{align}
    \label{gammaIV_lm}
    \gamma_{\ell m}^{I,V}(x_a,x_b,\hat{e}_a,\hat{e}_b) 
  &= \sum_{L_1 M_1 L_2 M_2} 
       J_{L_1} J^*_{L_2} Y_{L_1 M_1}(\hat{e}_a) Y^*_{L_2 M_2}(\hat{e}_b)
     \sum_{\ell_e m_e} \mathbb{D} \cdot \mathbb{E}^{I,V}_{\ell_e m_e} 
    \left\langle
    \begin{matrix}
        L_1 && L_2  && \ell_e   && \ell \\
        M_1 && M_2 && 0\; m_e  && 0\;  m
    \end{matrix}
    \right\rangle 
    \,, \\
    \label{gammaQU_lm}
 \gamma_{\ell m}^{Q\pm iU}(x_a,x_b,\hat{e}_a,\hat{e}_b) 
  &= \sum_{L_1 M_1 L_2 M_2} 
       J_{L_1} J^*_{L_2} Y_{L_1 M_1}(\hat{e}_a) Y^*_{L_2 M_2}(\hat{e}_b)
     \sum_{\ell_e m_e} \mathbb{D} \cdot \mathbb{E}^{Q\pm iU}_{\ell_e m_e} 
    \left\langle
    \begin{matrix}
        L_1 && L_2  && \ell_e   && \ell \\
        M_1 && M_2 && \mp4\; m_e  && \pm4\;  m
    \end{matrix}   
     \right\rangle 
    \,.
\end{align}
\ew
In Eqs.~(\ref{gammaIV_lm}) and (\ref{gammaQU_lm}), we have replaced the antenna pattern functions by
\be
\mathbb{D} \cdot \mathbb{E} \equiv  {}^{ijkl}\mathbb{D}_{ab} \mathbb{E}_{ijkl}\,,
\ee
and introduced a shorthand notation for the integral of a product of four spherical harmonics:
\bw
\be
    \left\langle
    \begin{matrix}
        L_1 && L_2 && l_1  &&  l_2 \\
        M_1 && M_2 && s_1\;m_1  && s_2\;m_2 
    \end{matrix}
    \right\rangle
    \equiv
    \int \d \hat{k} \; Y_{L_1 M_1}^*(\hat{k}) Y_{L_2 M_2}(\hat{k}) \;
     {}_{s_1}\!Y_{l_1 m_1}(\hat{k}) \; {}_{s_2}\!Y_{l_2 m_2}(\hat{k}) \,.
\ee
\ew
In Appendix~\ref{sec:Wigner3j}, we show that this integral can be expressed as a product of Wigner-3j symbols:
\bw
\begin{align}
 \left\langle
    \begin{matrix}
        L_1 && L_2 && \ell_e  &&  \ell \\
        M_1 && M_2 && s\;m_e  && -s\;m
    \end{matrix}
    \right\rangle 
=& \sum_{LM}\;
 (-1)^{M_1}\sqrt{\frac{(2L_1+1)(2L_2+1)(2L+1)}{4\pi}}
    \begin{pmatrix}
        L_1 && L_2  &&  L \\
        0 && 0  &&  0 
    \end{pmatrix}
    \begin{pmatrix}
        L_1 && L_2  &&  L \\
      -M_1 && M_2  &&  M 
    \end{pmatrix}\; \times\nonumber \\
 &   (-1)^M   \sqrt{\frac{(2L+1)(2\ell_e+1)(2\ell+1)}{4\pi}}
    \begin{pmatrix}
        L && \ell_e  &&  \ell \\
        0 && -s  &&  s
    \end{pmatrix}
    \begin{pmatrix}
        L && \ell_e  &&  \ell \\
       -M && m_e  &&  m
    \end{pmatrix} \,.
    \label{4product}
\end{align}
\ew
Eqs.~(\ref{gammaIV_lm}) and (\ref{gammaQU_lm}) are the most general ORFs for a pair of Galactic pulsars 
$a$ and $b$, respectively at distances $D_a$ and $D_b$ from the Earth. 

\section{Overlap Reduction Functions in the Computational Frame}

Let us choose the so-called computational frame such that pulsar $a$ is located at the north pole of the Earth 
while pulsar $b$ is stayed on the $\phi=0$ meridian. Their coordinates are then given by
\be
\hat{e}_a=(0,0,1),\quad \hat{e}_b=(\sin\zeta,0,\cos\zeta)\,
\ee
where $\zeta$ is their separation angle. In this coordinate system, the multipole moments of the antenna pattern functions denoted by $\mathbb{D}_0 \cdot \mathbb{E}$ are listed in Appendix~\ref{sec:DE}. Furthermore, we have
\be
Y_{L_1 M_1}(\hat{e}_a)= \sqrt{\frac{2L_1+1}{4\pi}} \delta_{M_1 0},\;\;
Y^*_{L_2 M_2}(\hat{e}_b)=Y^*_{L_2 M_2}(\zeta,0).
\ee
Thus, the integral of the product~(\ref{4product}) simplifies to
\bw
\begin{align}
 \left\langle
    \begin{matrix}
        L_1 && L_2 && \ell_e  &&  \ell \\
        M_1 && M_2 && s\;m_e  && -s\;m
    \end{matrix}
    \right\rangle 
 =&  \left\langle
    \begin{matrix}
        L_1 && L_2 && \ell_e  &&  \ell \\
        0 && M_2 && s\;m_e  && -s\;m
    \end{matrix}
    \right\rangle 
    \nonumber \\
=& \sum_{L}\;
 (-1)^{M_2}\sqrt{\frac{(2L_1+1)(2L_2+1)(2L+1)}{4\pi}}
    \begin{pmatrix}
        L_1 && L_2  &&  L \\
        0 && 0  &&  0 
    \end{pmatrix}
    \begin{pmatrix}
        L_1 && L_2  &&  L \\
        0 && M_2  &&  -M_2 
    \end{pmatrix}\; \times\nonumber \\
 &    \sqrt{\frac{(2L+1)(2\ell_e+1)(2\ell+1)}{4\pi}}
    \begin{pmatrix}
        L && \ell_e  &&  \ell \\
        0 && -s  &&  s
    \end{pmatrix}
    \begin{pmatrix}
        L && \ell_e  &&  \ell \\
       M_2 && m_e  &&  m
    \end{pmatrix} \,.
\end{align}
\ew

Hence, Eqs.~(\ref{gammaIV_lm}) and (\ref{gammaQU_lm}) become
\bw
\begin{align}
    \label{0gammaIV_lm}
    \gamma_{\ell m}^{I,V}(\zeta) 
  &= \sum_{L_1 L_2 M_2}   J_{L_1} J^*_{L_2}
      \sqrt{\frac{2L_1+1}{4\pi}} Y^*_{L_2 M_2}(\zeta,0)
     \sum_{\ell_e m_e} \mathbb{D}_0 \cdot \mathbb{E}^{I,V}_{\ell_e m_e} 
    \left\langle
    \begin{matrix}
        L_1 && L_2  && \ell_e   && \ell \\
        0  && M_2 && 0\; m_e  && 0\;  m
    \end{matrix}
    \right\rangle 
    \,, \\
    \label{0gammaQU_lm}
 \gamma_{\ell m}^{Q\pm iU}(\zeta) 
  &= \sum_{L_1 L_2 M_2}   J_{L_1} J^*_{L_2}
      \sqrt{\frac{2L_1+1}{4\pi}}  Y^*_{L_2 M_2}(\zeta,0)
     \sum_{\ell_e m_e} \mathbb{D}_0 \cdot \mathbb{E}^{Q\pm iU}_{\ell_e m_e} 
    \left\langle
    \begin{matrix}
        L_1 && L_2  && \ell_e   && \ell \\
        0 && M_2 && \mp4\; m_e  && \pm4\;  m
    \end{matrix}   
     \right\rangle 
    \,.
\end{align}
\ew

In Appendix~\ref{sec:DE}, we have found explicit analytic forms for the antenna pattern functions in the computational frame, 
$ \mathbb{D}_0 \cdot \mathbb{E}^{I,V}_{\ell_e m_e}$ and $\mathbb{D}_0 \cdot \mathbb{E}^{Q\pm iU}_{\ell_e m_e}$.
Every nonvanishing antenna pattern function is equal to the associated Legendre polynomial $P_2^{m_e} (\cos\zeta)$ times a real constant. 
Using Eq.~(\ref{Yconjugate}) and the conjugate relations for the antenna pattern functions in Appendix~\ref{sec:DE}, it is straightforward to show that the four ORFs in the computational frame have the conjugate relations:
\begin{align}
\gamma_{\ell -m}^{I}=&(-1)^{m} \gamma_{\ell m}^{I}\,,
\label{Iconjugate}\\
\gamma_{\ell -m}^{V}=&(-1)^{m+1}\gamma_{\ell m}^{V}\,,
\label{Vconjugate}\\
\gamma_{\ell -m}^{Q\pm iU}=&(-1)^{m} \gamma_{\ell m}^{Q\mp iU}\,.
\label{QUconjugate}
\end{align}
Note that the four ORFs are in general complex functions of $\zeta$. They become real when only the Earth term is considered~\cite{mingar13,kato16}. Our method includes the effects of the pulsar term that give rise to nonzero imaginary parts.

\subsection{Unpolarized Isotropic Case}

\begin{figure}[t]
\centering
\includegraphics[width=0.48\textwidth]{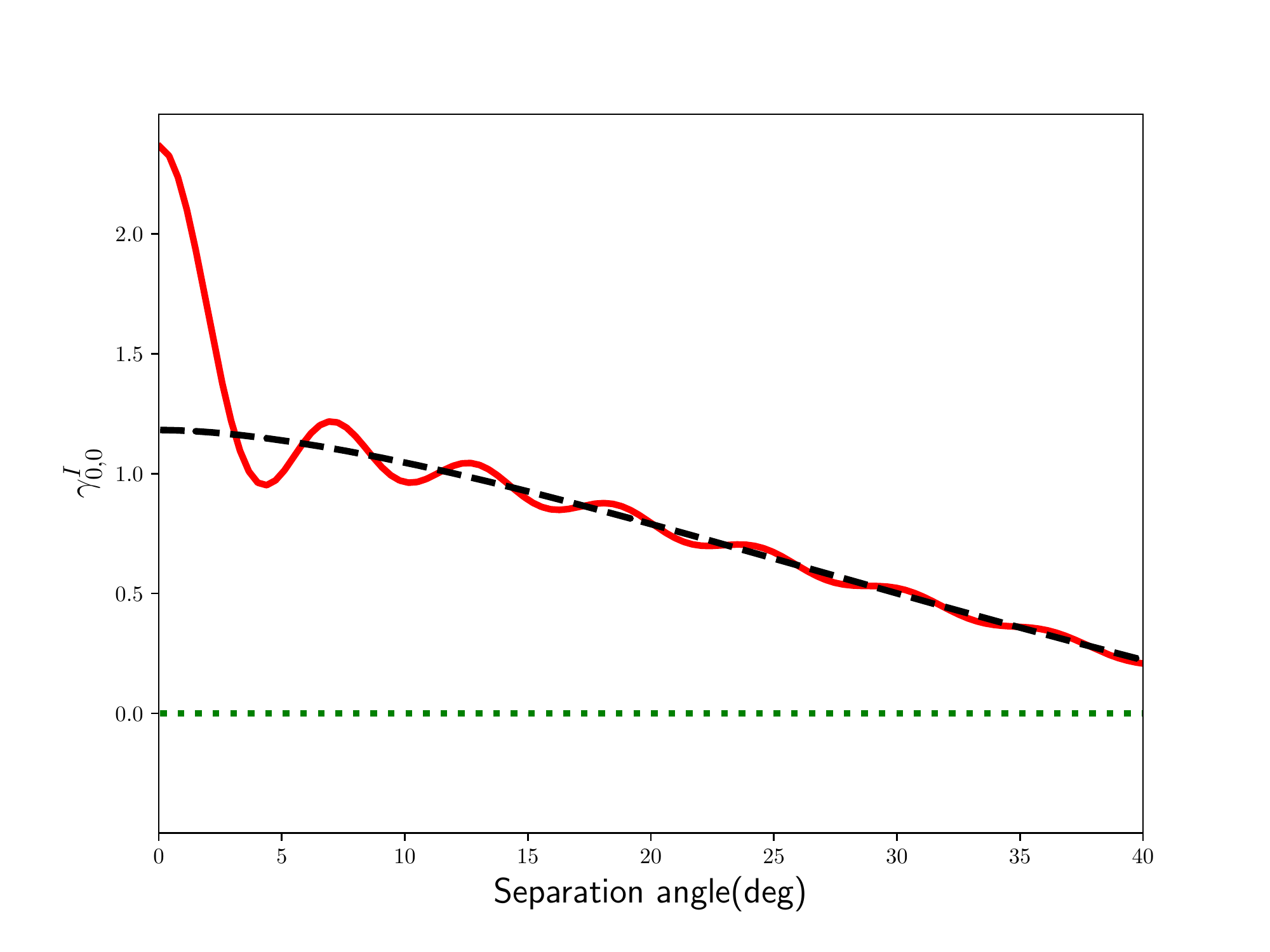}
\includegraphics[width=0.48\textwidth]{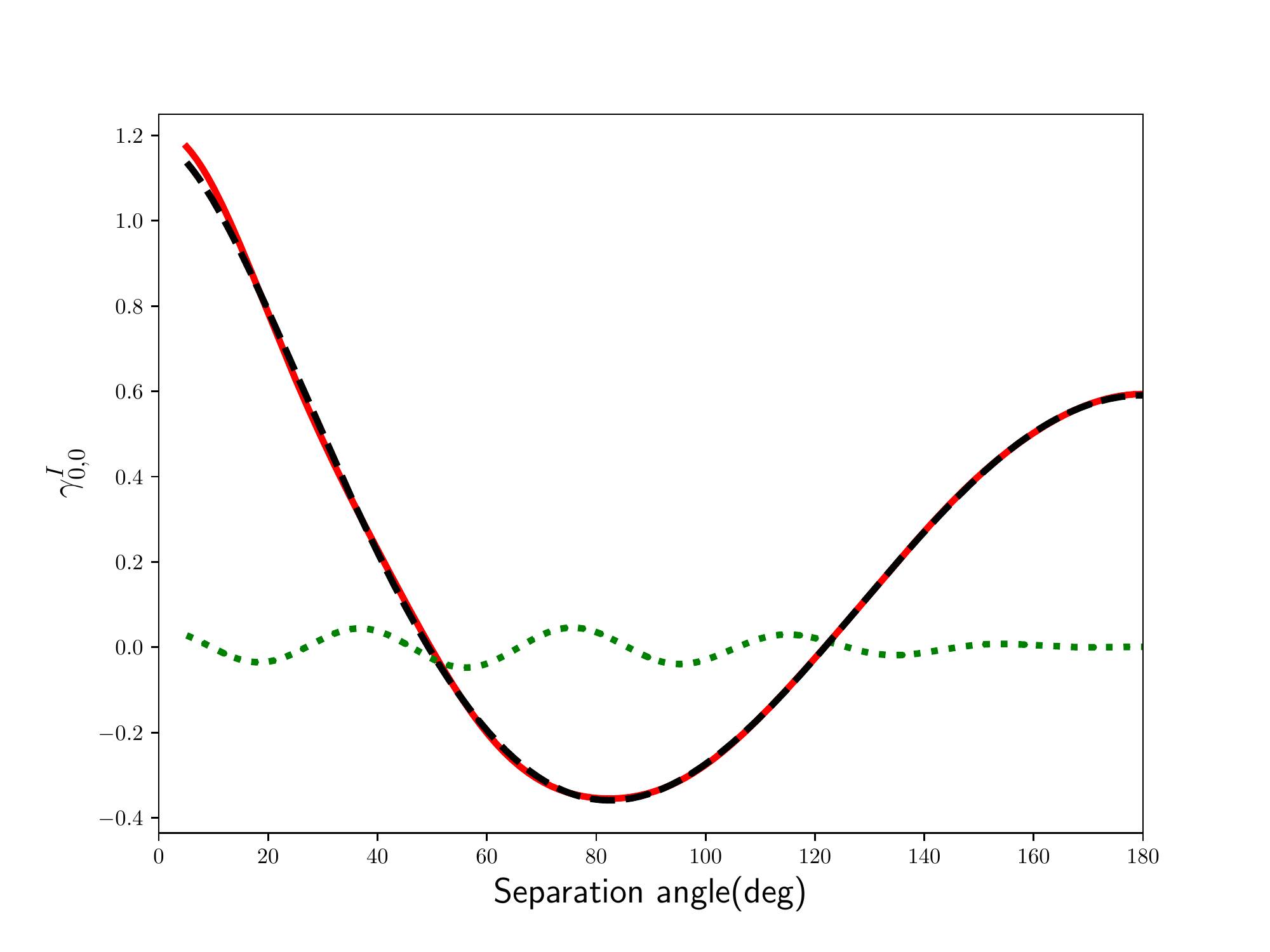}
\caption[Coordinate]{The right panel shows $\gamma_{00}^I(\zeta)$, while the left panel shows the power at small angular separation. We have used $x_a=x_b=20\pi$. The dashed curve is the Hellings and Downs curve. The real part of 
$\gamma_{00}^I(\zeta)$ (solid curves) overlaps with the Hellings and Downs curve at large angular separation, while the imaginary part of $\gamma_{00}^I(\zeta)$ (dotted curves) receives a small contribution from the pulsar term.}
\label{fig:gammaI00}
\index{figures}
\end{figure}

In the case of isotropic and unpolarized SGWB, the only relevant ORF is the $\gamma_{00}^{I}$, which can be calculated from Eq.~(\ref{0gammaIV_lm}) as
\bw
\begin{align}
   \gamma_{00}^I(\zeta) 
  =& \sum_{L_1 L_2 M_2} 
      J_{L_1} J^*_{L_2} \sqrt{\frac{2L_1+1}{4\pi}} Y^*_{L_2 M_2}(\zeta,0)
     \sum_{\ell_e m_e} \mathbb{D}_0 \cdot \mathbb{E}^I_{\ell_e m_e} 
    \left\langle
    \begin{matrix}
        L_1 && L_2  && \ell_e   && 0 \\
        0  && M_2 && 0\; m_e  && 0\;  0
    \end{matrix}
    \right\rangle 
     \nonumber \\
   =& \sum_{L_1 L_2 \ell_e m_e}  J_{L_1} J^*_{L_2}
      (-1)^{m_e} \frac{1}{4\pi}  Y_{L_2 m_e}(\zeta,0)\;
      \mathbb{D}_0 \cdot \mathbb{E}^I_{\ell_e m_e} \times \nonumber \\
     &\sqrt{\frac{(2L_1+1)^2(2L_2+1)(2\ell_e+1)}{4\pi}}
    \begin{pmatrix}
        L_1 && L_2  &&  \ell_e \\
        0 && 0  &&  0 
    \end{pmatrix}
      \begin{pmatrix}
        L_1 && L_2  &&  \ell_e \\
       0  && -m_e  &&  m_e
    \end{pmatrix} \,.
\end{align}
\ew
We have numerically computed $\gamma_{00}^I(\zeta)$ using $x_a=x_b=20\pi$ and summing $L_2$ up to $80$. Note that for each value of $L_2$, the range of $L_1$ is determined by the triangular condition in the Wigner-3j symbol. Figure~\ref{fig:gammaI00} shows $\gamma_{00}^I(\zeta)$ plotted against the separation angle $\zeta$. At large angular separation the real part of $\gamma_{00}^I(\zeta)$ reproduces the Hellings and Downs curve
for the quadrupolar interpulsar correlations~\cite{downs},
which has considered the Earth term only, while the pulsar term contributes to a relatively small imaginary part. 
$\gamma_{00}^I(\zeta)$ gains power from the pulsar term and begins to deviate from the Hellings and Downs curve at small angular separation. The autocorrelation $\gamma_{00}^I(\zeta=0^\circ)$ has power two times larger than the Hellings and Downs curve. This small-angle behavior is consistent with the results by considering the corrections from the pulsar term~\cite{mingar14}.

\subsection{Circularly Polarized Isotropic Case}
\label{CPI}

In the case of isotropic and circularly polarized SGWB, the other relevant ORF is the $\gamma_{00}^{V}$, which can be calculated from Eq.~(\ref{0gammaIV_lm}) as
\bw
\begin{align}
   \gamma_{00}^V(\zeta) 
  =& \sum_{L_1 L_2 M_2}  J_{L_1} J^*_{L_2}
       \sqrt{\frac{2L_1+1}{4\pi}} Y^*_{L_2 M_2}(\zeta,0)
     \sum_{\ell_e m_e} \mathbb{D}_0 \cdot \mathbb{E}^V_{\ell_e m_e} 
    \left\langle
    \begin{matrix}
        L_1 && L_2  && \ell_e   && 0 \\
        0  && M_2 && 0\; m_e  && 0\;  0
    \end{matrix}
    \right\rangle 
     \nonumber \\
    =& \sum_{L_1 L_2 \ell_e m_e} J_{L_1} J^*_{L_2}
      (-1)^{m_e} \frac{1}{4\pi}  Y_{L_2 m_e}(\zeta,0)\;
      \mathbb{D}_0 \cdot \mathbb{E}^V_{\ell_e m_e} \times \nonumber \\
     & \sqrt{\frac{(2L_1+1)^2(2L_2+1)(2\ell_e+1)}{4\pi}}
    \begin{pmatrix}
        L_1 && L_2  &&  \ell_e \\
        0 && 0  &&  0 
    \end{pmatrix}
      \begin{pmatrix}
        L_1 && L_2  &&  \ell_e \\
       0  && -m_e  &&  m_e
    \end{pmatrix} \,.  
\end{align}
\ew
We will show that $\gamma_{00}^V(\zeta)=0$ as follows. There is no contribution to the summation from terms with $m_e=0$ because $\mathbb{D}_0 \cdot \mathbb{E}^V_{\ell_e 0}=0$. Furthermore, since we have 
$\mathbb{D}_0 \cdot \mathbb{E}^V_{\ell_e -m_e}=(-1)^{m_e+1} \mathbb{D}_0 \cdot \mathbb{E}^V_{\ell_e m_e}$ 
and the property,
\be
 \begin{pmatrix}
        l_1 && l_2  &&  l_3 \\
       -m_1  && -m_2  && - m_3
  \end{pmatrix} 
  = (-1)^{l_1 + l_2 +  l_3}
  \begin{pmatrix}
        l_1 && l_2  &&  l_3 \\
       m_1  && m_2  &&  m_3
  \end{pmatrix} \,,
\ee
which is zero when $m_1=m_2=m_3=0$ unless $l_1 + l_2 +  l_3$ is an even integer, the contribution from terms with positive values of $m_e$ exactly cancels that from terms with negative values of $m_e$, 
making the overall summation vanish.

\subsection{Higher Multipole Moments}
\bw

\begin{figure}[ht!]
\centering
\includegraphics[width=0.48\textwidth]{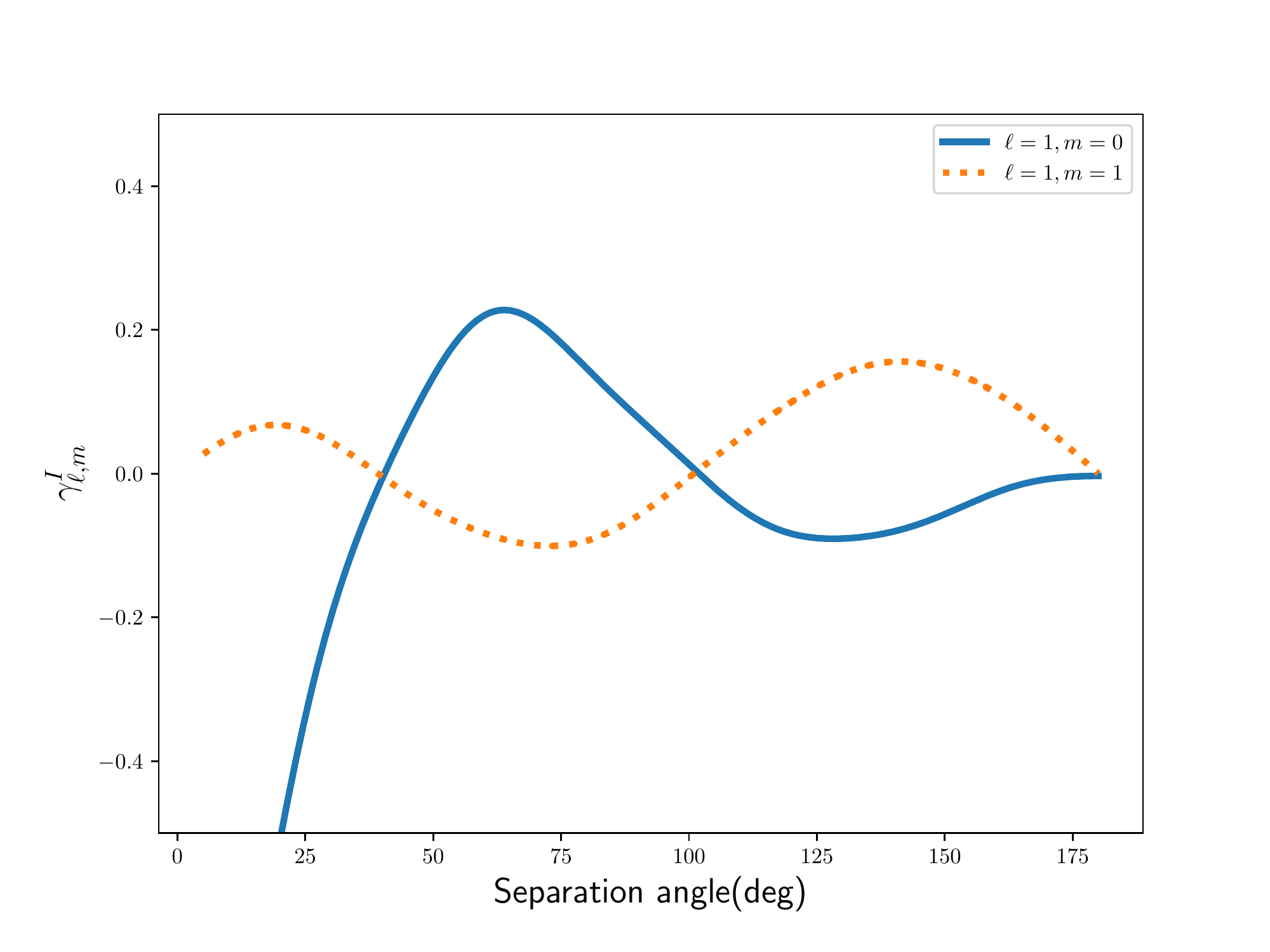}
\includegraphics[width=0.48\textwidth]{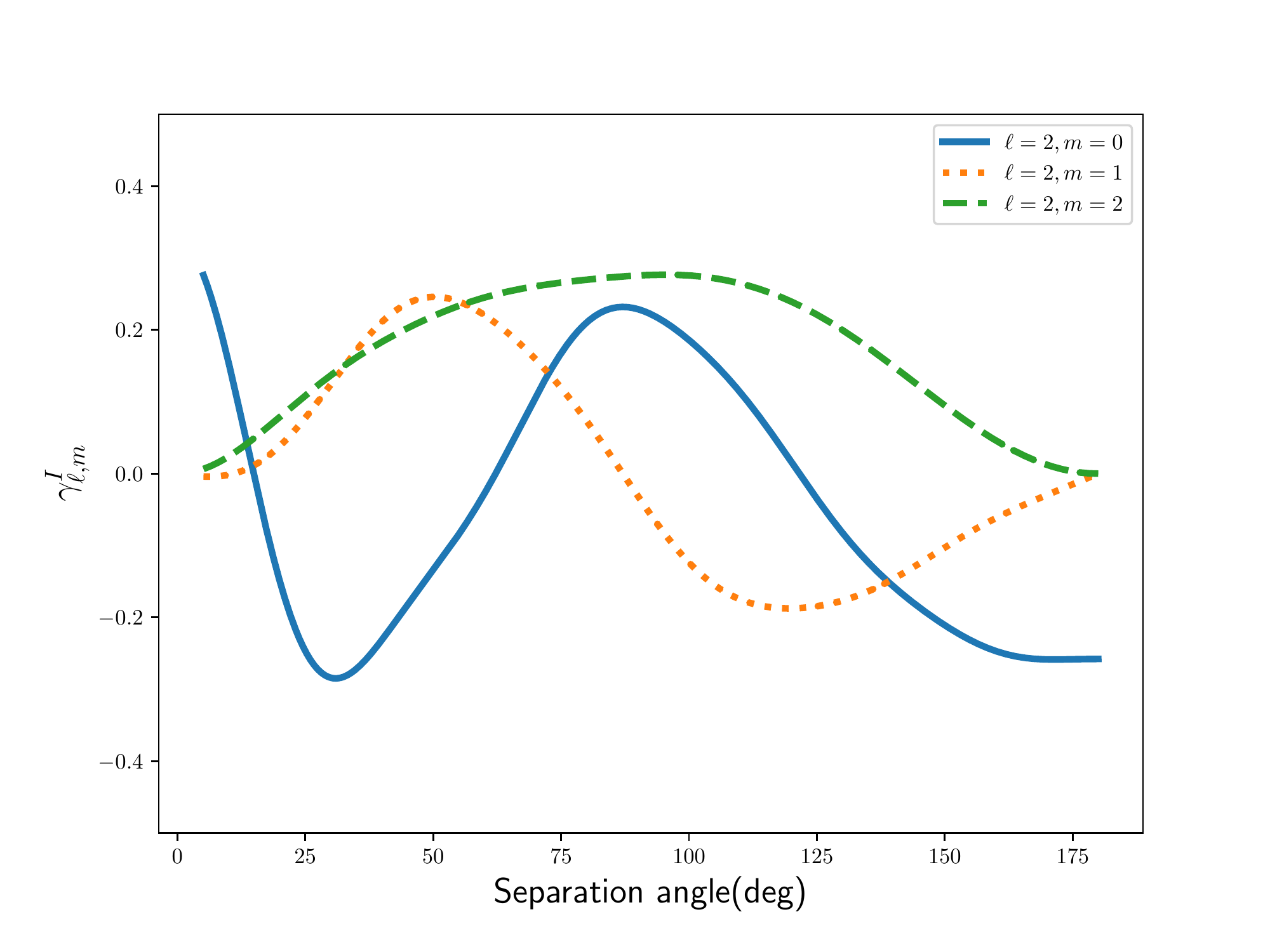}
\caption[Coordinate]{$\gamma_{\ell m}^I(\zeta)$ for $\ell=1,2$ and $m\ge 0$, with $x_a=x_b=20\pi$. The multipoles with $m<0$ are given by the relation~(\ref{Iconjugate}), $\gamma_{\ell -m}^{I}=(-1)^{m} \gamma_{\ell m}^{I}$.}
\label{fig:gammaIlm}
\index{figures}
\end{figure}
\begin{figure}[ht!]
\centering
\includegraphics[width=0.48\textwidth]{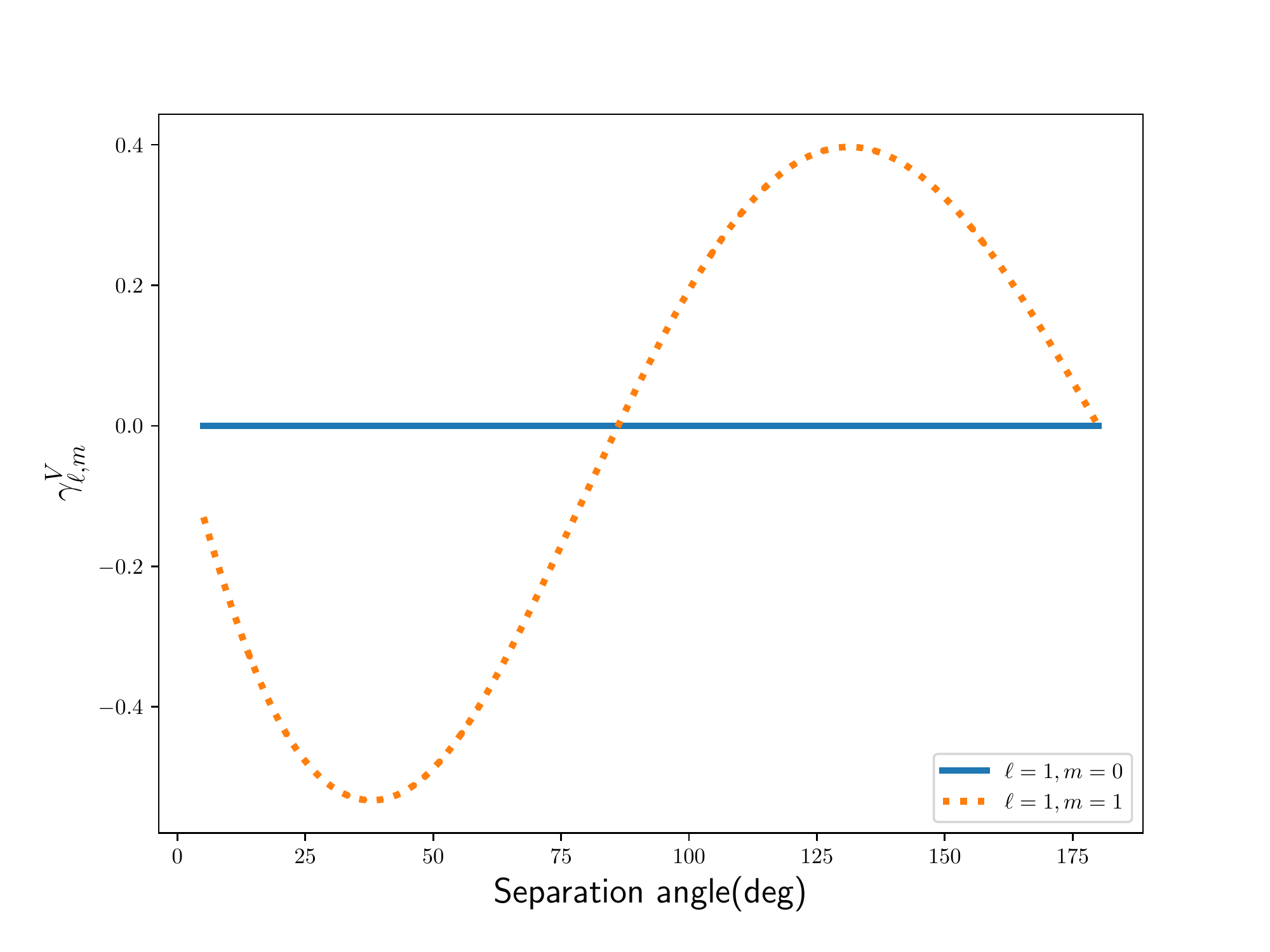}
\includegraphics[width=0.48\textwidth]{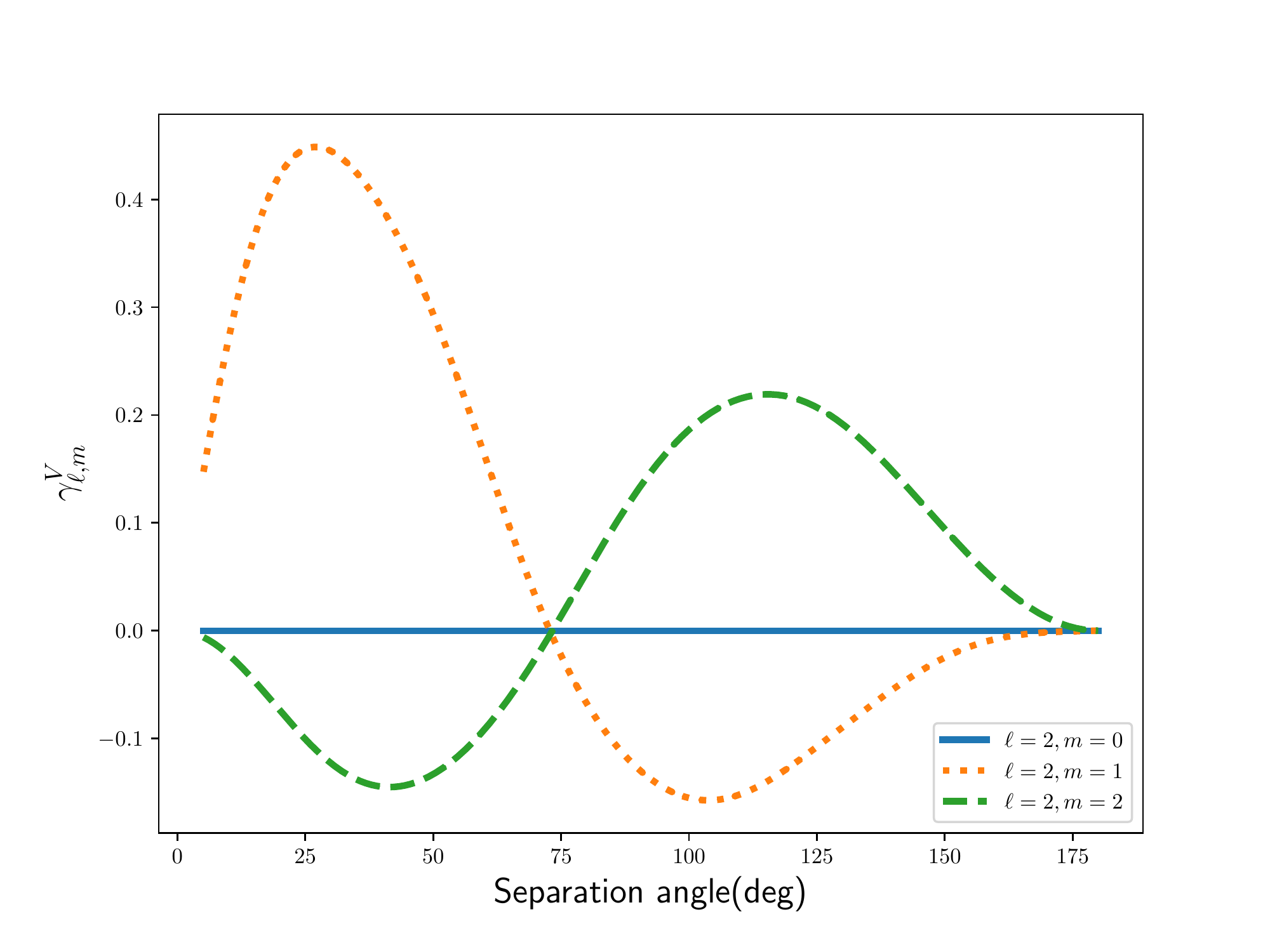}
\caption[Coordinate]{$\gamma_{\ell m}^V(\zeta)$ for $\ell=1,2$ and $m\ge 0$, with $x_a=x_b=20\pi$. The multipoles with $m<0$ are given by the relation~(\ref{Vconjugate}), $\gamma_{\ell -m}^{V}=(-1)^{m+1} \gamma_{\ell m}^{V}$.}
\label{fig:gammaVlm}
\index{figures}
\end{figure}
\begin{figure}[ht!]
\centering
\includegraphics[width=0.48\textwidth]{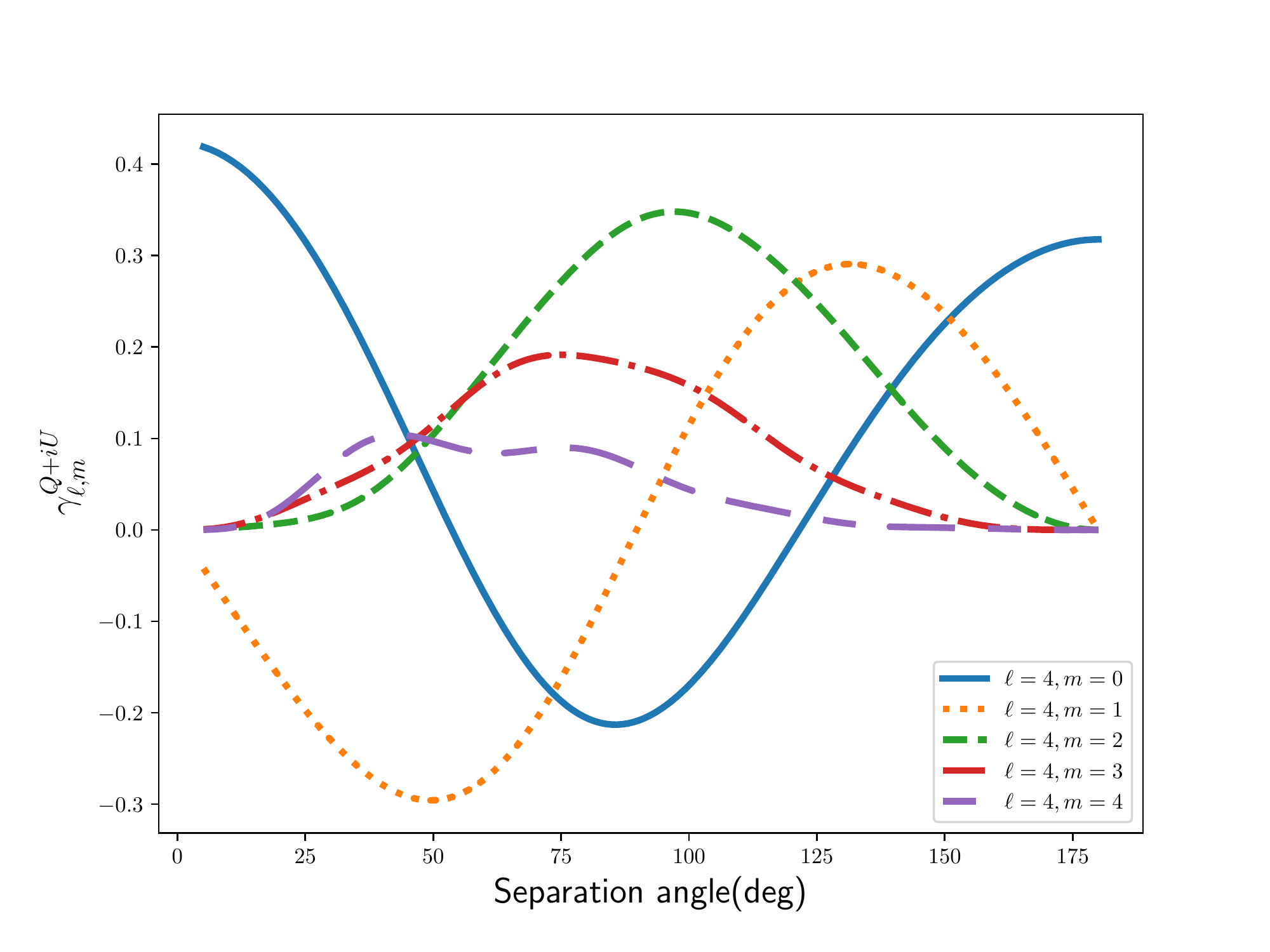}
\includegraphics[width=0.48\textwidth]{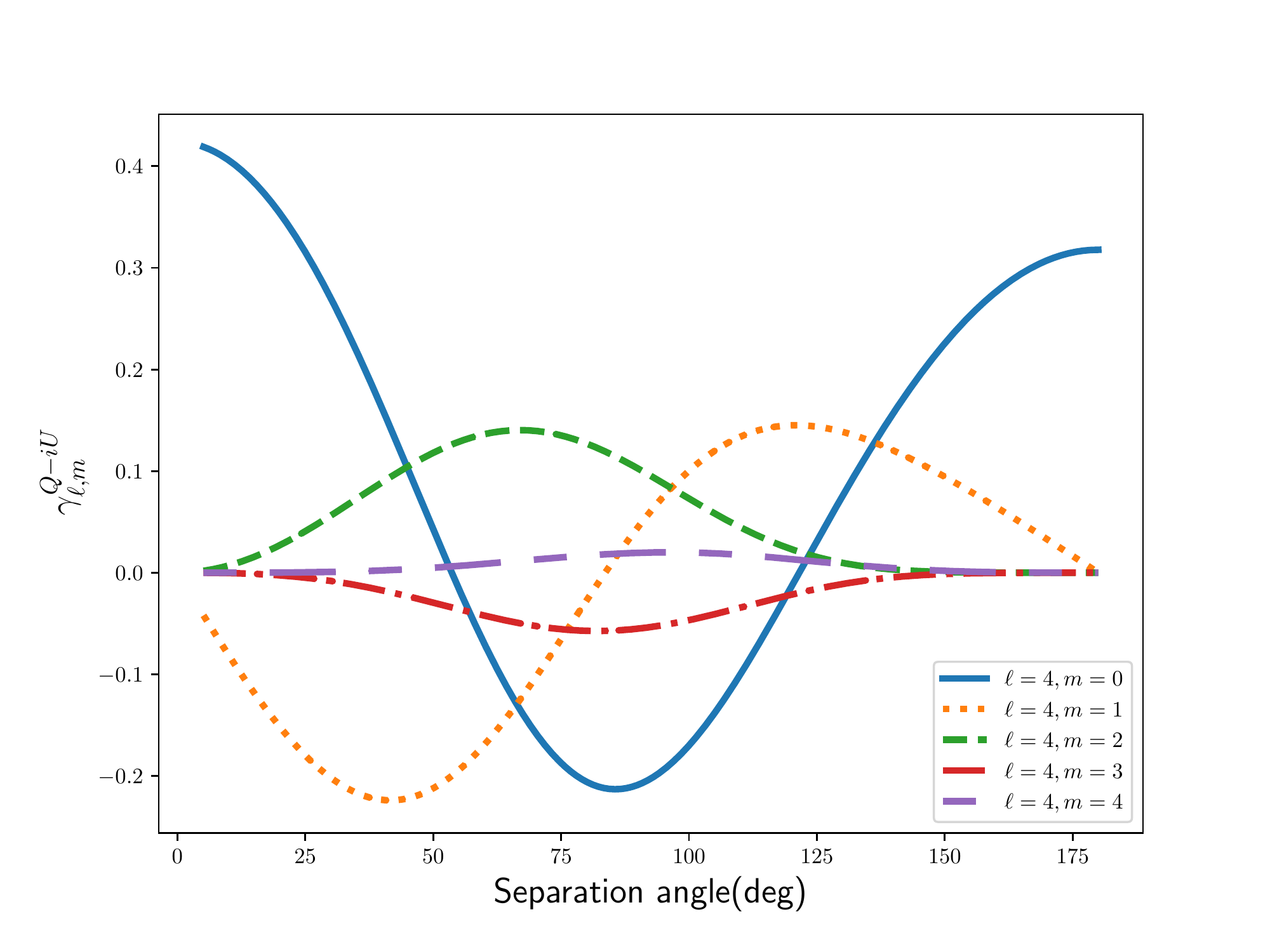}
\caption[Coordinate]{$\gamma_{\ell m}^{Q\pm iU}(\zeta)$ for $\ell=4$ and $m\ge 0$, with $x_a=x_b=20\pi$. 
The multipoles with $m<0$ are given by the relation~(\ref{QUconjugate}),
$\gamma_{\ell -m}^{Q\pm iU}=(-1)^{m} \gamma_{\ell m}^{Q\mp iU}$.}
\label{fig:gammaQUlm}
\index{figures}
\end{figure}

\ew
We have numerically computed $\gamma_{\ell m}^I(\zeta)$ using $x_a=x_b=20\pi$ and summing $L_2$ up to $8$ for 
$\ell=1,2$, as shown in Fig.~\ref{fig:gammaIlm}. Note that we have plotted the ORFs for $\zeta>5^\circ$. To obtain the small-angle resolution that shows the effects from the pulsar term, we would need to increase the maximum value of $L_2$. Similarly, Fig.~\ref{fig:gammaVlm} shows $\gamma_{\ell m}^V(\zeta)$ for $\ell=1,2$. Note that $\gamma_{\ell 0}^V(\zeta)=0$, which can be shown following the same reasoning as in Sec.~\ref{CPI}. These ORF harmonics reproduce those found in Refs.~\cite{mingar13,kato16} using the Shapiro time delay with the Earth term only. Figure~\ref{fig:gammaQUlm} shows our new results for $\gamma_{\ell m}^{Q\pm iU}(\zeta)$ for $\ell=4$. In these figures, we have omitted the imaginary parts that are small compared to the real parts.
\bw

\ew

\section{Overlap Reduction Functions in the Celestial Coordinates}

We can readily get the ORFs in the celestial coordinates from those in the computational frame through a three-dimensional rotation~\cite{chu21}. Let pulsar $a$ be located at the polar angles $\hat{e}_a(\theta,\phi)$, and the angle between its longitude and the great circle connecting the pulsar pair $a$-$b$ be $\alpha$. Then, Eqs.~(\ref{gammaIV_lm}) and (\ref{gammaQU_lm}) are connected to Eqs.~(\ref{0gammaIV_lm}) and (\ref{0gammaQU_lm}) by
\begin{align}
    \gamma_{\ell m}^{I,V}(\hat{e}_a,\hat{e}_b) 
  &= \sum_{m'} D^{\ell}_{m' m}(-\alpha,-\theta,-\phi) \gamma_{\ell m'}^{I,V}(\zeta)   \,, \\
 \gamma_{\ell m}^{Q\pm iU}(\hat{e}_a,\hat{e}_b) 
  &=  \sum_{m'} D^{\ell}_{m' m}(-\alpha,-\theta,-\phi) \gamma_{\ell m'}^{Q\pm iU}(\zeta)   \,,
\end{align}
where the three-dimensional rotation is represented by the Wigner-$D$ matrix, which is closely related to the spin-weighted spherical harmonics:
\begin{align}
    \label{eq:DtoSWSH}
    D^{\ell}_{s m}(\alpha,\beta,\gamma)
    = \sqrt{\frac{4\pi}{2\ell+1}} {}_{-s}Y_{\ell m}(-\beta,-\gamma) e^{-i s \alpha} \,.
\end{align}

\section{Conclusion}

We have developed a numerical scheme to compute the ORFs for the Stokes parameters of an anisotropic stochastic gravitational wave background in pulsar-timing-array observation. We have used the Sachs-Wolfe line-of-sight integral to construct the time-residual correlation between a pair of Galactic millisecond pulsars on the sky. The integral incorporates simultaneously the effects of the pulsar term that contribute power to the ORFs at small angular separation of the pulsar pair. The ORFs for the linear polarization are given for the first time. Based on a spherical harmonic analysis, our method allows us to compute the ORF multipoles order by order as well as improving their resolution at small angular separation in a controllable manner. The method can be readily applied to compute the ORFs for a Galactic pulsar pair at different distances from the Earth. Furthermore, the Sachs-Wolfe line-of-sight integral can be used to consider extragalactic pulsars at high redshifts.

\begin{acknowledgments}
This work was supported in part by the Ministry of Science and Technology (MOST) of Taiwan, Republic of China, under
Grants No. MOST 109-2112-M-032-006 (G.C.L.) and No. MOST 109-2112-M-001-003 (K.W.N.).
\end{acknowledgments}

\appendix
\bw
\section{Spin-Weighted Spherical Harmonics}
\label{sec:spinweight}
The explicit form of the spin-weighted spherical harmonics that we use is
\begin{align}
{}_{s}Y_{\ell m}(\theta,\phi) = 
    (-1)^{s+m}
    e^{im\phi}
    \sqrt{\frac{(2\ell+1)}{(4\pi)}\frac{(\ell+m)!(\ell-m)!}{(\ell+s)!(\ell-s)!}}
    \sin^{2\ell}\!\left(\frac{\theta}{2}\right)
    \sum_{r}
    \binom{\ell-s}{r} \binom{\ell+s}{r+s-m}
    (-1)^{\ell-r-s}
    \cot^{2r+s-m}\!\left(\frac{\theta}{2}\right) \,.
\end{align}
\ew
When $s=0$, it reduces to the ordinary spherical harmonics,
\be
Y_{\ell m}(\hat{n})=\sqrt{\frac{(2\ell+1)}{(4\pi)}\frac{(\ell-m)!}{(\ell+m)!}}P^m_\ell(\cos \theta) e^{i m \phi} \,.
\ee

Spin-weighted spherical harmonics satisfy the orthogonal relation,
\be
    \int_{S^2} \d{\hat{n}}\; {}_{s}Y^*_{\ell m}(\hat{n}){}_{s}Y_{\ell' m'}(\hat{n})
    = \delta_{\ell \ell'} \delta_{m m'} \,,
\ee
and the completeness relation,
\begin{align}
    \sum_{\ell m} {}_{s}Y^*_{\ell m}(\hat{n}){}_{s}Y_{\ell m}(\hat{n}')
    =& \delta(\hat{n}-\hat{n}') \nonumber \\
    =& \delta(\phi-\phi')\delta(\cos\theta-\cos\theta')\,.
\end{align}
Its complex conjugate is
\be
{}_{s}Y^*_{\ell m}(\hat{n}) =  (-1)^{s+m} {}_{-s}Y_{\ell -m}(\hat{n}) \,,
\label{Yconjugate}
\ee
and its parity is given by
\be
{}_{s}Y_{\ell m}(-\hat{n}) \equiv {}_{s}Y_{\ell m}(\pi-\theta,\phi+\pi)=(-1)^{\ell} {}_{-s}Y_{\ell m}(\hat{n}) \,.
\label{Yparity}
\ee
Also, we have the spherical wave expansion:
\be
\label{eq:swexpansion}
e^{i\vec{k}\cdot\vec{r}} = 4\pi \suml \summ i^\ell j_\ell(kr) Y_{\ell m}^*(\hat{k}) Y_{\ell m}(\hat{r}) \,,
\ee
where $j_\ell(x)$ is the spherical Bessel function. 


\bw
\section{Multipole Moments of Polarization Tensor}
\label{sec:polartensor}
\ew
\subsection{${}_{ijkl} \mathbb{E}_{\ell m}^{I}$}
The only nonzero coefficients are $\ell=0,2,4$ cases for ${}_{ijkl} \mathbb{E}_{\ell m}^{I}$, which are symmetric under exchanging between $i \leftrightarrow j$, $k \leftrightarrow l$, and ${ij} \leftrightarrow {kl}$. In addition, they satisfy the relation $\mathbb{E}_{\ell -m}^I=(-1)^m \mathbb{E}^{I*}_{\ell m}$.
\begin{align*}
    \frac{16}{15}
    \sqrt{\pi}
  &= {}_{xxxx}\mathbb{E}_{00} 
   = {}_{yyyy}\mathbb{E}_{00}
   = {}_{zzzz}\mathbb{E}_{00}
  \\
    -\frac{8}{15}
    \sqrt{\pi}
  &= {}_{xxyy}\mathbb{E}_{00} 
   = {}_{xxzz}\mathbb{E}_{00} 
   = {}_{yyzz}\mathbb{E}_{00} 
\end{align*}
\begin{align*}
    \frac{16}{21}
    \sqrt{\frac{\pi}{5}}
  &= {}_{xxxx}\mathbb{E}_{20} 
   = {}_{yyyy}\mathbb{E}_{20} 
   = {}_{xxzz}\mathbb{E}_{20}
   = {}_{yyzz}\mathbb{E}_{20}
   \\
    -\frac{32}{21}
    \sqrt{\frac{\pi}{5}}
  &= {}_{zzzz}\mathbb{E}_{20} 
   = {}_{xxyy}\mathbb{E}_{20} 
\end{align*}
\begin{align*}
    \frac{4}{7}
    \sqrt{\frac{2\pi}{15}}
  &= {}_{xxxz}\mathbb{E}_{21} 
   = {}_{xzzz}\mathbb{E}_{21} 
   \\
    -\frac{4i}{7}
    \sqrt{\frac{2\pi}{15}}
  &= {}_{yyyz}\mathbb{E}_{21} 
   = {}_{yzzz}\mathbb{E}_{21} 
   \\
    \frac{2}{7}
    \sqrt{\frac{6\pi}{5}}
  &= {}_{xxyz}\mathbb{E}_{21} 
   = {}_{xyyz}\mathbb{E}_{21} 
\end{align*}
\begin{align*}
    \frac{8}{7}
    \sqrt{\frac{2\pi}{15}}
  &= {}_{xxzz}\mathbb{E}_{22} 
   = {}_{yyyy}\mathbb{E}_{22} 
   \\
    -\frac{8}{7}
    \sqrt{\frac{2\pi}{15}}
  &= {}_{xxxx}\mathbb{E}_{22} 
   = {}_{yyzz}\mathbb{E}_{22} 
   \\
    -\frac{8i}{7}
    \sqrt{\frac{2\pi}{15}}
  &= {}_{xyzz}\mathbb{E}_{22}
   \\
    \frac{4i}{7}
    \sqrt{\frac{2\pi}{15}}
  &= {}_{xxxy}\mathbb{E}_{22} 
   = {}_{xyyy}\mathbb{E}_{22} 
\end{align*}
\begin{align*}
    \frac{2}{35}
    \sqrt{\pi}
  &= {}_{xxxx}\mathbb{E}_{40} 
   = {}_{yyyy}\mathbb{E}_{40} 
   \\
    \frac{2}{105}
    \sqrt{\pi}
  &= {}_{xxyy}\mathbb{E}_{40} 
   \\
    -\frac{8}{105}
    \sqrt{\pi}
  &= {}_{xxzz}\mathbb{E}_{40} 
   = {}_{yyzz}\mathbb{E}_{40} 
   \\
    \frac{16}{105}
    \sqrt{\pi}
  &= {}_{zzzz}\mathbb{E}_{40} 
\end{align*}
\begin{align*}
    \frac{1}{7}
    \sqrt{\frac{\pi}{5}}
  &= {}_{xxxz}\mathbb{E}_{41} 
   \\
    -\frac{i}{7}
    \sqrt{\frac{\pi}{5}}
  &= {}_{yyyz}\mathbb{E}_{41} 
   \\
    \frac{1}{21}
    \sqrt{\frac{\pi}{5}}
  &= {}_{xyyz}\mathbb{E}_{41} 
   \\
    -\frac{i}{21}
    \sqrt{\frac{\pi}{5}}
  &= {}_{xxyz}\mathbb{E}_{41} 
   \\
    -\frac{4}{21}
    \sqrt{\frac{\pi}{5}}
  &= {}_{xzzz}\mathbb{E}_{41} 
   \\
    \frac{4i}{21}
    \sqrt{\frac{\pi}{5}}
  &= {}_{yzzz}\mathbb{E}_{41} 
\end{align*}
\begin{align*}
    \frac{2}{21}
    \sqrt{\frac{2\pi}{5}}
  &= {}_{yyyy}\mathbb{E}_{42} 
   = {}_{xxzz}\mathbb{E}_{42} 
   \\
    -\frac{2}{21}
    \sqrt{\frac{2\pi}{5}}
  &= {}_{xxxx}\mathbb{E}_{42} 
   = {}_{yyzz}\mathbb{E}_{42} 
   \\
    \frac{2i}{21}
    \sqrt{\frac{2\pi}{5}}
  &= {}_{xyzz}\mathbb{E}_{42} 
   \\
    \frac{i}{21}
    \sqrt{\frac{2\pi}{5}}
  &= {}_{xxxy}\mathbb{E}_{42} 
   = {}_{xyyy}\mathbb{E}_{42} 
\end{align*}
\begin{align*}
    \frac{1}{3}
    \sqrt{\frac{\pi}{35}}
  &= {}_{xyyz}\mathbb{E}_{43} 
   \\
    -\frac{1}{3}
    \sqrt{\frac{\pi}{35}}
  &= {}_{xxxz}\mathbb{E}_{43}
   \\
    -\frac{i}{3}
    \sqrt{\frac{\pi}{35}}
  &= {}_{yyyz}\mathbb{E}_{43} 
   \\
    \frac{i}{3}
    \sqrt{\frac{\pi}{35}}
  &= {}_{xxyz}\mathbb{E}_{43} 
\end{align*}
\begin{align*}
    \frac{1}{3}
    \sqrt{\frac{\pi}{35}}
  &= {}_{xyyz}\mathbb{E}_{43} 
   \\
    -\frac{1}{3}
    \sqrt{\frac{\pi}{35}}
  &= {}_{xxxz}\mathbb{E}_{43}
   \\
    -\frac{i}{3}
    \sqrt{\frac{\pi}{35}}
  &= {}_{yyyz}\mathbb{E}_{43} 
   \\
    \frac{i}{3}
    \sqrt{\frac{\pi}{35}}
  &= {}_{xxyz}\mathbb{E}_{43} 
\end{align*}
\begin{align*}
    \frac{1}{3}
    \sqrt{\frac{2\pi}{35}}
  &= {}_{xxxx}\mathbb{E}_{44} 
   = {}_{yyyy}\mathbb{E}_{44} 
   \\
    -\frac{i}{3}
    \sqrt{\frac{2\pi}{35}}
  &= {}_{xxxy}\mathbb{E}_{44}
   \\
    -\frac{1}{3}
    \sqrt{\frac{2\pi}{35}}
  &= {}_{xxyy}\mathbb{E}_{44} 
   \\
    \frac{i}{3}
    \sqrt{\frac{2\pi}{35}}
  &= {}_{xyyy}\mathbb{E}_{44} 
\end{align*}

\subsection{${}_{ijkl} \mathbb{E}_{\ell m}^{V}$}
The only nonzero coefficients are $\ell=1,3$ cases for ${}_{ijkl} \mathbb{E}_{\ell m}^{V}$, which are symmetric under exchanging between $i \leftrightarrow j$ and $k \leftrightarrow l$, and are antisymmetric under exchanging ${ij} \leftrightarrow {kl}$. In addition, they satisfy the relation $\mathbb{E}^V_{\ell -m}=(-1)^{m+1} \mathbb{E}^{V*}_{\ell m}$.  
\begin{align*}
    -\frac{8i}{5}
    \sqrt{\frac{\pi}{3}}
  &= {}_{xxxy}\mathbb{E}_{10} 
   = {}_{xyyy}\mathbb{E}_{10} 
\end{align*}
\begin{align*}
    -\frac{4}{5}
    \sqrt{\frac{2\pi}{3}}
  &= {}_{xxxz}\mathbb{E}_{11} 
   = {}_{xzzz}\mathbb{E}_{11} 
   \\
    \frac{4i}{5}
    \sqrt{\frac{2\pi}{3}}
  &= {}_{yyyz}\mathbb{E}_{11} 
   = {}_{yzzz}\mathbb{E}_{11} 
   \\
    -\frac{2}{5}
    \sqrt{\frac{2\pi}{3}}
  &= {}_{xyyz}\mathbb{E}_{11} 
\end{align*}
\begin{align*}
    -\frac{2i}{5}
    \sqrt{\frac{\pi}{7}}
  &= {}_{xxxy}\mathbb{E}_{30} 
   = {}_{xyyy}\mathbb{E}_{30} 
\end{align*}
\begin{align*}
    -\frac{1}{5}
    \sqrt{\frac{\pi}{21}}
  &= {}_{xxxz}\mathbb{E}_{31} 
   \\
    -i
    \sqrt{\frac{\pi}{21}}
  &= {}_{xxyz}\mathbb{E}_{31} 
   \\
    -\frac{1}{5}
    \sqrt{\frac{3\pi}{7}}
  &= {}_{xyyz}\mathbb{E}_{31} 
   \\
    \frac{4}{5}
    \sqrt{\frac{\pi}{21}}
  &= {}_{xzzz}\mathbb{E}_{31} 
   \\
    \frac{i}{5}
    \sqrt{\frac{\pi}{21}}
  &= {}_{yyyz}\mathbb{E}_{31} 
   \\
    -\frac{4i}{5}
    \sqrt{\frac{\pi}{21}}
  &= {}_{yzzz}\mathbb{E}_{31} 
\end{align*}
\begin{align*}
    i
    \sqrt{\frac{2\pi}{105}}
  &= {}_{xxxy}\mathbb{E}_{32} 
  \\
    2
    \sqrt{\frac{2\pi}{105}}
  &= {}_{xxyy}\mathbb{E}_{32} 
   = {}_{yyzz}\mathbb{E}_{32} 
  \\
    -2
    \sqrt{\frac{2\pi}{105}}
  &= {}_{xxzz}\mathbb{E}_{32} 
  \\
    -i
    \sqrt{\frac{2\pi}{105}}
  &= {}_{xyyy}\mathbb{E}_{32} 
  \\
    2i
    \sqrt{\frac{2\pi}{105}}
  &= {}_{xyzz}\mathbb{E}_{32} 
\end{align*}
\begin{align*}
    \sqrt{\frac{\pi}{35}}
  &= {}_{xxxz}\mathbb{E}_{33} 
  \\
    -i
    \sqrt{\frac{\pi}{35}}
  &= {}_{xxyz}\mathbb{E}_{33} 
  \\
    -
    \sqrt{\frac{\pi}{35}}
  &= {}_{xyyz}\mathbb{E}_{33} 
  \\
    i
    \sqrt{\frac{\pi}{35}}
  &= {}_{yyyz}\mathbb{E}_{33} 
\end{align*}

\subsection{${}_{ijkl} \mathbb{E}_{\ell m}^{Q\pm iU}$}
The only nonzero coefficients are $\ell=4$ cases for ${}_{ijkl} \mathbb{E}_{\ell m}^{Q\pm iU}$, which are symmetric under exchanging between $i \leftrightarrow j$, $k \leftrightarrow l$, and ${ij} \leftrightarrow {kl}$. In addition, they satisfy the relation $\mathbb{E}_{\ell -m}^{Q\pm iU}=(-1)^m \mathbb{E}_{\ell m}^{Q\mp iU*}$. 
Here, specifically we have $\mathbb{E}_{\ell m}^{Q+iU}=\mathbb{E}_{\ell m}^{Q-iU}$.
\begin{align*}
    \sqrt{\frac{2\pi}{35}}
  &= {}_{xxxx}\mathbb{E}_{40} 
   = {}_{yyyy}\mathbb{E}_{40}
  \\
    \frac{1}{3}
    \sqrt{\frac{2\pi}{35}}
 &= {}_{xxyy}\mathbb{E}_{40} 
  \\
    -\frac{4}{3}
    \sqrt{\frac{2\pi}{35}}
 &= {}_{xxzz}\mathbb{E}_{40} 
  = {}_{yyzz}\mathbb{E}_{40}
  \\
    \frac{8}{3}
    \sqrt{\frac{2\pi}{35}}
 &= {}_{zzzz}\mathbb{E}_{40} 
\end{align*}

\begin{align*}
    \sqrt{\frac{\pi}{14}}
  &= {}_{xxxz}\mathbb{E}_{41} 
  \\
    -\frac{i}{3}
    \sqrt{\frac{\pi}{14}}
  &= {}_{xxyz}\mathbb{E}_{41}
  \\
    \frac{1}{3}
    \sqrt{\frac{\pi}{14}}
  &= {}_{xyyz}\mathbb{E}_{41} 
  \\
    -\frac{2}{3}
    \sqrt{\frac{2\pi}{7}}
  &= {}_{xzzz}\mathbb{E}_{41} 
  \\
    -i
    \sqrt{\frac{\pi}{14}}
  &= {}_{yyyz}\mathbb{E}_{41}
  \\
    \frac{2i}{3}
    \sqrt{\frac{\pi}{7}}
  &= {}_{yzzz}\mathbb{E}_{41} 
\end{align*}

\begin{align*}
    \frac{2}{3}
    \sqrt{\frac{\pi}{7}}
  &= {}_{xxzz}\mathbb{E}_{42}
   = {}_{yyyy}\mathbb{E}_{42}
   \\
    -\frac{2}{3}
    \sqrt{\frac{\pi}{7}}
  &= {}_{xxxx}\mathbb{E}_{42} 
   = {}_{yyzz}\mathbb{E}_{42} 
  \\
    \frac{i}{3}
    \sqrt{\frac{\pi}{7}}
  &= {}_{xxxy}\mathbb{E}_{42}
   = {}_{xyyy}\mathbb{E}_{42} 
\end{align*}

\begin{align*}
    \frac{1}{3}
    \sqrt{\frac{\pi}{2}}
  &= {}_{xyyz}\mathbb{E}_{43}
   = {}_{xxxz}\mathbb{E}_{43}
  \\
    \frac{i}{3}
    \sqrt{\frac{\pi}{2}}
  &= {}_{xxyz}\mathbb{E}_{43}
   = {}_{yyyz}\mathbb{E}_{43} 
\end{align*}

\begin{align*}
    \frac{1}{3}
    \sqrt{\pi}
  &= {}_{xxxx}\mathbb{E}_{44}
   = {}_{yyyy}\mathbb{E}_{44}
   \\
    -\frac{1}{3}
    \sqrt{\pi}
  &= {}_{xxyy}\mathbb{E}_{44}
  \\
    \frac{i}{3}
    \sqrt{\pi}
  &= {}_{xyyy}\mathbb{E}_{44}
  \\
    -\frac{i}{3}
    \sqrt{\pi}
  &= {}_{xxxy}\mathbb{E}_{44} 
\end{align*}

\section{Wigner-3j Symbols} 
\label{sec:Wigner3j}

We first introduce the integral of a product of three spherical harmonics,
\bw
\be
    \int \d \hat{k} \;
    Y_{LM}(\hat{k})\; {}_{s_1}\!Y_{l_1 m_1}(\hat{k}) \; {}_{s_2}\!Y_{l_2 m_2}(\hat{k})=
    \sqrt{\frac{(2L+1)(2l_1+1)(2l_2+1)}{4\pi}}
    \begin{pmatrix}
        L && l_1  &&  l_2 \\
        0 && -s_1  &&  -s_2 
    \end{pmatrix}
    \begin{pmatrix}
        L && l_1  &&  l_2 \\
       M && m_1  &&  m_2 
    \end{pmatrix} \,,
\label{eq:threeJ}
\ee
\ew
which involves two Wigner-3j symbols representing the coupling coefficients between different spherical harmonics~\cite{book:Varshalovich}.  
It is worth noting that the properties of the Wigner-3j symbols in Eq.~(\ref{eq:threeJ}) imply that $s_1+s_2=0$. In addition, $L$, $l_1$, and $l_2$ have to satisfy the triangular condition, i.e., $l_1 + l_2 \ge L \ge |l_1- l_2|$, while $M+m_1+m_2=0$. 

To evaluate the integral of a product of four spherical harmonics, we expand the product of two spherical harmonics as
\be
Y_{L_1 M_1}^*(\hat{k}) Y_{L_2 M_2}(\hat{k}) = \sum_{LM} c_{L_1 M_1 L_2 M_2}^{LM} Y^*_{LM}(\hat{k})\,.
\ee
Using Eq.~(\ref{eq:threeJ}), we obtain
\bw
\be
 c_{L_1 M_1 L_2 M_2}^{LM}= (-1)^{M_1}\sqrt{\frac{(2L_1+1)(2L_2+1)(2L+1)}{4\pi}}
    \begin{pmatrix}
        L_1 && L_2  &&  L \\
        0 && 0  &&  0 
    \end{pmatrix}
    \begin{pmatrix}
        L_1 && L_2  &&  L \\
      -M_1 && M_2  &&  M 
    \end{pmatrix} \,.
\ee
\ew
Then,
\bw
\begin{align}
 &\int \d \hat{k} \; Y_{L_1 M_1}^*(\hat{k}) Y_{L_2 M_2}(\hat{k}) \;
     {}_{s_1}\!Y_{l_1 m_1}(\hat{k}) \; {}_{s_2}\!Y_{l_2 m_2}(\hat{k}) \nonumber \\
=& \sum_{LM}\;
 (-1)^{M_1}\sqrt{\frac{(2L_1+1)(2L_2+1)(2L+1)}{4\pi}}
    \begin{pmatrix}
        L_1 && L_2  &&  L \\
        0 && 0  &&  0 
    \end{pmatrix}
    \begin{pmatrix}
        L_1 && L_2  &&  L \\
      -M_1 && M_2  &&  M 
    \end{pmatrix}\; \times\nonumber \\
 &   (-1)^M   \sqrt{\frac{(2L+1)(2l_1+1)(2l_2+1)}{4\pi}}
    \begin{pmatrix}
        L && l_1  &&  l_2 \\
        0 && -s_1  &&  -s_2 
    \end{pmatrix}
    \begin{pmatrix}
        L && l_1  &&  l_2 \\
       -M && m_1  &&  m_2 
    \end{pmatrix} \,.
\end{align}
\ew

\section{Antenna Pattern Functions}
\label{sec:DE}
In most literature, the inner product between the detector tensor and the polarization basis tensor is referred to as the antenna pattern function. 

\subsection{$\mathbb{DE}^I$}
\label{sec:DEI}
For the Stokes-I parts, the only nonvanishing $\mathbb{D}_0(\zeta) \cdot \mathbb{E}^{I}_{\ell m}$ are 15 components with $\ell=0,2,4$, which satisfy $\mathbb{DE}_{\ell -m}=(-1)^m \mathbb{DE}^*_{\ell m}$.

\begin{align*}
    \mathbb{DE}^I_{00}
    &=\frac{4}{15} \sqrt{\pi} \left(1+3\cos(2\zeta)\right)=\frac{16}{15} \sqrt{\pi} \,P^0_2(\cos\zeta)
    \\
    \mathbb{DE}^I_{20}
    &=-\frac{8}{21} \sqrt{\frac{\pi}{5}} \left(1+3\cos(2\zeta)\right)=-\frac{32}{21} \sqrt{\frac{\pi}{5}} \,P^0_2(\cos\zeta)
     \\
    \mathbb{DE}^I_{21}
    &=\frac{8}{7} \sqrt{\frac{2\pi}{15}} \cos(\zeta) \sin(\zeta)=-\frac{8}{21} \sqrt{\frac{2\pi}{15}}  \,P^1_2(\cos\zeta)
     \\
    \mathbb{DE}^I_{22}
    &=\frac{8}{7} \sqrt{\frac{2\pi}{15}} \sin^2(\zeta)= \frac{8}{21} \sqrt{\frac{2\pi}{15}} \,P^2_2(\cos\zeta)
     \\
    \mathbb{DE}^I_{40}
    &=\frac{4}{105} \sqrt{\pi} \left(1+3\cos(2\zeta)\right)= \frac{16}{105} \sqrt{\pi} \,P^0_2(\cos\zeta)
      \\
    \mathbb{DE}^I_{41}
    &=-\frac{8}{21} \sqrt{\frac{\pi}{5}} \cos(\zeta)\sin(\zeta)=\frac{8}{63} \sqrt{\frac{\pi}{5}}  \,P^1_2(\cos\zeta)
      \\
    \mathbb{DE}^I_{42}
    &=\frac{2}{21} \sqrt{\frac{2\pi}{5}} \sin^2(\zeta)= \frac{2}{63} \sqrt{\frac{2\pi}{5}} \,P^2_2(\cos\zeta)
     \\
    \mathbb{DE}^I_{43}
    &=0
     \\
    \mathbb{DE}^I_{44}
    &=0
 \end{align*}

\subsection{$\mathbb{DE}^V$}
\label{sec:DEV}
For the Stokes-V parts that correspond to the circular polarized signal, the only nonvanishing $\mathbb{D}_0(\zeta) \cdot \mathbb{E}^{V}_{\ell m}$ are 10 components with $\ell=1,3$, which satisfy $\mathbb{DE}_{\ell -m}=(-1)^{m+1} \mathbb{DE}^*_{\ell m}$.

\begin{align*}
    \mathbb{DE}^V_{10}
    &=0 
      \\
    \mathbb{DE}^V_{11}
    &=\frac{8}{5} \sqrt{\frac{2\pi}{3}} \cos(\zeta)\sin(\zeta) = -\frac{8}{15} \sqrt{\frac{2\pi}{3}}\,P^1_2(\cos\zeta) 
    \\
    \mathbb{DE}^V_{30}
    &=0 
     \\
    \mathbb{DE}^V_{31}
    &=-\frac{8}{5} \sqrt{\frac{\pi}{21}} \cos(\zeta)\sin(\zeta)=\frac{8}{15} \sqrt{\frac{\pi}{21}}\,P^1_2(\cos\zeta)  
       \\
    \mathbb{DE}^V_{32}
    &=2 \sqrt{\frac{2\pi}{105}} \sin^2(\zeta)=\frac{2}{3} \sqrt{\frac{2\pi}{105}} \,P^2_2(\cos\zeta) 
       \\
    \mathbb{DE}^V_{33}
    &=0 
 \end{align*}

\subsection{$\mathbb{DE}^{Q\pm iU}$}
For the linear polarized signal, the only nonvanishing $\mathbb{D}_0(\zeta) \cdot \mathbb{E}^{Q\pm i U}_{\ell m}$ are 9 components with $\ell=4$, which satisfy $\mathbb{DE}_{\ell -m}=(-1)^m \mathbb{DE}^*_{\ell m}$.

\begin{align*}
    \mathbb{DE}^{Q\pm iU}_{40}
    &=\frac{2}{3} \sqrt{\frac{2\pi}{35}} \left(1+3\cos(2\zeta)\right)= \frac{8}{3} \sqrt{\frac{2\pi}{35}} \,P^0_2(\cos\zeta)
    \\
    \mathbb{DE}^{Q\pm iU}_{41}
    &=-\frac{4}{3} \sqrt{\frac{2\pi}{7}} \cos(\zeta)\sin(\zeta) =\frac{4}{9} \sqrt{\frac{2\pi}{7}} \,P^1_2(\cos\zeta)
    \\
    \mathbb{DE}^{Q\pm iU}_{42}
    &=\frac{2}{3} \sqrt{\frac{\pi}{7}} \sin^2(\zeta)=\frac{2}{9} \sqrt{\frac{\pi}{7}} \,P^2_2(\cos\zeta)
    \\
    \mathbb{DE}^{Q\pm iU}_{43}
    &=0
    \\
    \mathbb{DE}^{Q\pm iU}_{44}
    &=0
\end{align*}

\newcommand{\Authname}[2]{#2 #1} 
\newcommand{\etal}{{\it et al.}}
\newcommand{\LSC}{\Authname{Abbott}{B.~P.} \etal\, (LIGO Scientific Collaboration)}
\newcommand{\LVC}{\Authname{Abbott}{B.~P.} \etal\, (LIGO Scientific and Virgo Collaboration)}
\newcommand{\LVK}{\Authname{Abbott}{B.~P.} \etal\, (LIGO Scientific, Virgo and KAGRA Collaboration)}

\newcommand{\Title}[1]{}               

\newcommand{\arxiv}[1]{\href{http://arxiv.org/abs/#1}{{arXiv:}#1}}
\newcommand{\PRD}[3]{\href{https://doi.org/10.1103/PhysRevD.#1.#2}{{Phys. Rev. D} {\bf #1}, #2 (#3)}}
\newcommand{\PRL}[3]{\href{https://doi.org/10.1103/PhysRevLett.#1.#2}{{Phys. Rev. Lett.} {\bf #1}, #2 (#3)}}

\newcommand{\MNRAS}[4]{\href{https://doi.org/10.1093/mnras/#1.#4.#2}{{Mon. Not. R. Astron. Soc.} {\bf #1}, #2 (#3)}}
\newcommand{\CQGii}[5]{\href{https://doi.org/10.1088/0264-9381/#1/#4/#5}{{Class. Quant. Grav.} {\bf #1}, #2 (#3)}}
\newcommand{\CQG}[4]{\href{https://doi.org/10.1088/1361-6382/#4}{{Class. Quant. Grav.} {\bf #1}, #2 (#3)}}
\newcommand{\JCAP}[3]{\href{https://doi.org/10.1088/1475-7516/#3/#1/#2}{{J. Cosmol. Astropart. Phys.} #1 (#3) #2}}

\newcommand{\LRR}[4]{\href{https://doi.org/10.1007/#4}{{Liv. Rev. Rel.} {\bf #1}, #2 (#3)}}
\newcommand{\ApJ}[4]{\href{https://doi.org/#4}{{Astrophys. J.} {\bf #1}, #2 (#3)}}


\raggedright

\end{document}